\documentclass[twocolumn,preprintnumbers,amsmath,amssymb]{revtex4}
\usepackage{graphicx}% Include figure files
\usepackage{dcolumn}% Align table columns on decimal point
\usepackage{bm}% bold math
\usepackage{amsmath}

\begin{document}

%\preprint{APS/123-QED}

\title {First Penning-trap mass measurement in the  millisecond half-life range: the exotic halo nucleus $^{11}$Li}

\author{M.~Smith$^{*,1,2}$}
\author{M.~Brodeur$^{1,2}$}
\author{T.~Brunner$^{1,3}$}
\author{S. Ettenauer$^{1,2}$}

\author{A~Lapierre$^{1}$}
\author{R.~Ringle$^{1}$}
\author{V. L. ~Ryjkov$^{1}$}

\author{F.~Ames$^{1}$}

\author{P.~Bricault$^{1}$}
\author{G. W. F.~Drake$^{4}$}
\author{P.~Delheij$^{1}$}
\author{D~Lunney$^{1,5}$}
\author{J.~Dilling$^{1,2}$}

\affiliation{$^{1}$TRIUMF, 4004 Wesbrook Mall, Vancouver, British Columbia, Canada}
\affiliation{$^{2}$Physics and Astronomy, University of British Columbia, 6224 Agricultural Road, Vancouver BC, Canada}

\affiliation{$^{3}$Technische Universit\"at M\"unchen, E12, James Franck Strasse, Garching, Germany}
\affiliation{$^{4}$Department of Physics, University of Windsor, Windsor, Ontario, Canada}
\affiliation{$^{5}$CSNSM/CNRS/IN2P3, Universite« de Paris-Sud, F-91405, Orsay, France}
\date{\today}% It is always \today, today,
             %  but any date may be explicitly specified

\begin{abstract}

In this letter, we report a new mass for $^{11}$Li using the trapping experiment
TITAN at TRIUMF's ISAC facility.   This is by far the shortest-lived nuclide, $t_{1/2} = 8.8 \, \rm{ms}$,
for which a mass measurement has ever been performed with a Penning trap.
Combined with our mass measurements of $^{8,9}$Li we derive a new two-neutron separation energy
of 369.15(65) keV: a factor of seven more precise than the best
previous value.  This new value is a critical ingredient for the determination of
the halo charge radius from isotope-shift measurements.  We also report
results from state-of-the-art atomic-physics calculations using the new mass
and extract a new charge radius for $^{11}$Li.  This result is a remarkable
confluence of nuclear and atomic physics.
\end{abstract}

\pacs{21.10.Dr,21.10.Gv,21.45.-v,,29.30.Aj}
%21.10.Dr:  binding energies and masses
%21.10.Gv:   Nucleon distributions and halo features
%21.45.-v:  few-body systems
%27.20.+n:  6 < A < 19
%29.30.Aj:  charged-particle spectrometers:  electric and magnetic
%29.38.-c:  radioactive beams
\keywords{Penning Trap, Mass Spectrometer, Halo Nuclei, Charge Radius}%Use showkeys class option if keyword

\maketitle

Perched precariously on the brink of nuclear stability,
$^{11}$Li has the lowest two-neutron binding energy of all bound
nuclear systems~\cite{bau07}.  This gives rise to the exotic phenomenon
of a nuclear halo that has a wavefunction extending beyond
the range normally allowed for by the strong interaction.
The determination of the all-important $^{11}$Li binding energy,
from a direct mass measurement, is particularly challenging
due to its half-life of a mere 8.8 ms.  

The abnormal spatial properties of the
original, and most famous, halo nuclide $^{11}$Li  were discovered by Tanihata \textit{et al.} in
1985 \cite{Tani85}.  Since then examples of nuclei with one ($^{11}$Be), two
($^6$He, $^{14}$Be, $^{17}$B) and even four ($^8$He) neutron halos,  have also been found.  Nuclei with two neutron halos, such as $^{11}$Li, are especially interesting 
because they form three body systems where none of 
the composite two body systems are themselves 
stable. Such nuclei were dubbed Borromean by M. Zhukov after the heraldic symbol of the Borromeo family; three rings linked together in such a 
way that removal of any one ring leaves the other two disconnected.  

The halo phenomena arises because of a very low nucleon binding energy, typically hundreds of keV, more than an order of magnitude smaller than that of
stable nuclei.  Because halo nuclei are so weakly bound their properties provide stringent 
tests for nuclear models.  As such a large amount of work, both experimental and theoretical, has been carried out on the subject, see e.g. \cite{Jons04,Jens04} for reviews.   Interest in the archetypical halo nucleus $^{11}$Li has recently been rekindled due a number of recent experimental results including the nuclear charge radius, determined with precision laser spectroscopy by Sanchez \textit{et al.} \cite{San06} and the soft electric-dipole excitation, measured through invariant mass spectrometry by Nakamura \textit{et al.}
\cite{Naka06}; the results of both of which are dependent on the $^{11}$Li mass.  Reproduction of these results provides a real challenge to models of $^{11}$Li; a number of which themselves use the two neutron separation energy, $S_{2n}$, to to adjust the $^{9}$Li-$n$ interaction (see, for example \cite{Vin95,For02,Brid06,Myo07}).

Of the various techniques used for
mass measurements (see, for example, the review \cite{lun03}),
Penning traps have emerged as the most accurate balances for
weighing atomic nuclei, where for  stable ions relative uncertainies of about 10$^{-11}$ have been obtained~\cite{Rai04}.   High precision measurements on radioactive species are much more difficult due constraints which arise due to short half-lives.  Before this work the shortest lived isotope measured with this technique was $^{74}$Rb,   t$_{1/2}$ = 65 ms, to a precision of $\delta m / m = 6 \times 10^{-8}$~\cite{kel03}.

A precision Penning trap is formed by  the superposition of a harmonic electrostatic potential over  a homogeneous  magnetic field.    This results in a trapping potential in which an ion undergoes three independent harmonic eigenmotions: a longitudinal oscillation, $\nu_{z}$, due to the applied electric field, a fast radial cyclotron motion, $\nu_+$, due to the longitudinal magnetic field and a slow radial magnetron drifting, $\nu_-$, caused by the perpendicular magnetic and electric fields~\cite{gab86}.  It can be shown that the sum of the frequencies of these two radial motions is equal to that of the true cyclotron frequency of the ion:
\begin{equation}\label{cyc_freq}
\nu_c = \nu_+ + \nu_-  = \frac{1}{2 \pi} \frac{q}{m} B,
\end{equation}
where $q/m$ is the charge to mass ratio of the ion and $B$ is the magnetic field strength. 

\begin{figure}
\includegraphics[width=1\columnwidth]{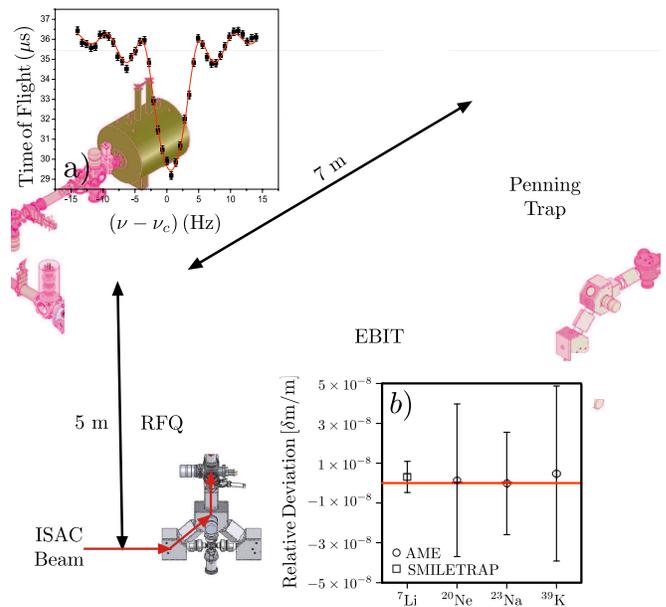}
\caption{ Layout of the TITAN Penning trap mass spectrometer.  The ISAC beam is first cooled and bunched in a gas-filled RFQ before being delivered to the next stage of the experiment.   In general ions are then sent to the electron beam ion trap (EBIT) which is used for charge state breeding. In this experiment due to the extremely short half-life of $^{11}$Li the EBIT was bypassed and the ions were injected directly into the precision Penning trap. a) A typical $^6$Li time of flight spectrum is shown with a fit of the theoretical line-shape, here $\nu_c$ = 9\,450\,927 Hz.  b) Results of measurements of a range of masses with respect to that of $^6$Li.  These are compared to literature values, from the AME~\cite{ame03} and SMILETRAP~\cite{nag06}.  \label{TITAN}
 }
\end{figure}

A measurement of the cyclotron frequency of an ion in a Penning trap can be made by the application of an azimuthal quadrupolar rf field of  frequency $\nu_{\rm{rf}}$.  Such a field causes a periodic conversion between the two radial motions, which is most efficient in resonance i.e. when $\nu_{\rm{rf}} = \nu_{\rm{c}}$.  For any given excitation time, $T_{\rm{rf}}$, the amplitude of the rf can be set such that a full conversion between magnetron and reduced cyclotron motions occurs only in resonance.  If an ion is injected into the trap in such a way that it has only an initial slow magnetron motion in the radial plane the application of the rf field at the resonant frequency acts so as to increase the radial energy of the ion up to a maximum achieved when the ion's motion is fully converted into pure reduced cyclotron.  The radial energy after the excitation as a function of $\nu_{\rm{rf}}$, has a characteristic shape with a large peak centred at $\nu_{\rm{rf}} = \nu_{\rm{c}}$, and a number of successively smaller peaks known as sidebands (see figure~\ref{TITAN}a).  A measurement of the radial kinetic energy of the ions can be made using the time of flight technique where the ions are extracted from the trap and guided electrostatically toward an MCP detector~\cite{graf80}.  During this process the ions  traverse an in-homogeneous magnetic field which has the effect of converting their radial kinetic energy into longitudinal kinetic energy.  Thus, the ion's time of flight between the trap and the detector is a minimum when the initial radial kinetic energy is a maximum i.e.\ when $\nu_{\rm{rf}} = \nu_{\rm{c}}$.  The width of the resonance curve, $\Delta \nu$, and ultimately the resolution of the measurement can be shown to be inversely proportional to the excitation time $\Delta \nu \approx 1 / T_{rf}$.  Hence, when extending this technique down to very short-lived isotopes it is important to reduce the amount of time required for the preparation of the ions to a minimum so as to maximise the rf excitation time.

To calculate the mass of an ion from its cyclotron frequency using equation (\ref{cyc_freq}) the magnetic field must be known. This is found experimentally by measuring the cyclotron frequency of a reference ion, $\nu_{ref}$, whose mass, $m_{ref}$, is well known.  The ratio of the reference frequency to that of the ion of interest, $r = \nu_{\rm{ref}} / \nu_{\rm{c}}$, can then be used to calculate the new mass as:
\begin{equation}\label{mass}
m =   r(m_{\rm{ref}} -m_{\rm{e}})+m_{\rm{e}},
\end{equation}
where $m_{\rm{e}}$ is the electron mass.

The experiments described were carried out at TRIUMF's ISAC facility.  The radioactive lithium beams were produced using a tantalum thin-foil target, bombarded by a 500-MeV, continuous proton beam of 70 $\mu \rm{A}$.   The resulting reaction products were surface ionised before being accelerated to a transport energy of  20  keV.  The isotopes of interest were selected by passing the beam through a two stage magnetic dipole separator.  $^{11}$Li was produced at a rate of approximately 3,000 ions/s. 

The TITAN spectrometer, shown in figure~\ref{TITAN}, consists of three ion traps~\cite{del06}, two of which were used to make the measurements described here.   The 20 keV ISAC beam was first electrostatically decelerated to a longitudinal kinetic energy of a few eV before injection into a gas filled radiofrequency quadrupole (RFQ) beam cooler.   In these devices, now in common use at radioactive beam facilities, the injected beam is thermalised via successive interactions with a room temperature gas, in this case molecular hydrogen (H$_2$).  The cooled ions are then collected in a longitudinal potential well before pulsed extraction. 

The cooled ions were then sent directly to a 3.7 T measurement Penning trap.  This is in contrast to other online Penning trap facilities where an extra step of ion preparation (which can include isobaric purification) is usually carried out in a separate Penning trap.  For these measurements this was unnecessary due to the contaminant free lithium beams.  Once trapped, ions are conventionally given an initial pure magnetron motion via the application of a dipole excitation. This excitation takes a fixed amount of time, typically on the order of a few milliseconds, thus reducing the time available for the measurement.  However, at TITAN an electromagnetic (Lorentz~\cite{rin07}) steering device was used to give the required magnetron obit. This acts on the ions as they are being injected into the trap avoiding an additional excitation period.  

\begin{figure}
\includegraphics[width=1\columnwidth]{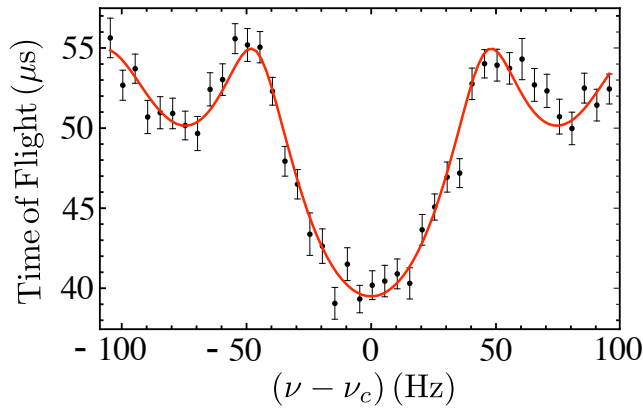}
\caption{ A typical $^{11}$Li resonance collected over 30 minutes, containing approximately 1000 ions.  Here $\nu_c$ = 5\,147\,555 Hz . The solid line is a fit of the theoretical curve~\cite{kon95} to the data. \label{11result}
 }
\end{figure}

A total of nine separate measurements of the $^{11}$Li cyclotron frequency, each taking approximately 30 minutes,  were made over a fourteen hour period.  An excitation time of two half-lifes (18 ms) was used.  Reference measurements were made using $^{6}$Li at one hour intervals.  Measurements of the masses of $^{8,9}$Li were also made, using a 48 ms excitation time.  The data were analysed following as closely as possible the well established procedure of the ISOLTRAP experiment~\cite{kel03}.  A typical $^{11}$Li resonance is shown in figure~\ref{11result}, the central frequency was found from a fit of the theoretical line shape (as illustrated)~\cite{kon95}.   The results for the frequency ratios for these lithium isotopes are shown in table~\ref{tab:meas}.   From these ratios new values for the mass excess of these isotopes were derived using the recent SMILETRAP measurement of the mass excess of  $^6$Li, $\Delta(^6$Li)  = 14086.882(37) keV ~\cite{nag06}.  The quoted values include a systematic error which takes into account both linear ($\delta m / m = 2 \times 10^{-9}$) and non-linear ($\delta m / m = 7 \times 10^{-9}$) drifts of the magnetic field which where added in quadrature to the statistical uncertainty.  The effects of deviations from the ideal electric and magnetic fields were also explicitly probed by measurement of a range of nuclei ($4<m<39 \,  \rm{u}$), with respect to $^6$Li, in all cases agreement, within error bars, was obtained between the the TITAN measurements and the literature values (see figure~\ref{TITAN}b).  An upper limit on these effects was then derived from the uncertainty in the TITAN measurements as $\delta m / m = 1.5 \times 10^{-9}$ per mass unit difference between the measured and reference ions (i.e. $7.5\times10^{-9}$ for $^{11}$Li).  This was added linearly into the final error budget.

\begin{figure}
\includegraphics[width=1\columnwidth]{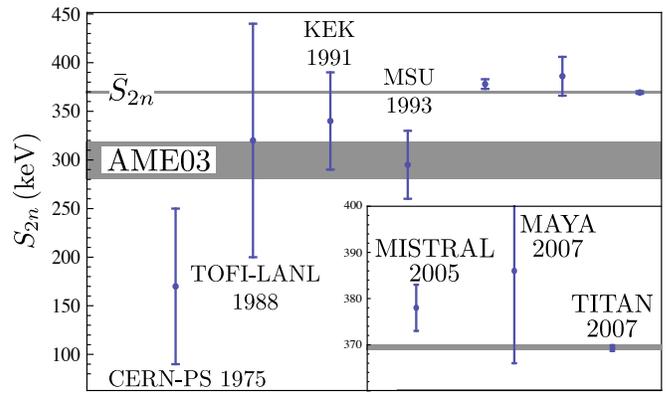}
\caption{$^{11}$Li two-neutron separation energies derived from
previous mass measurements: CERN-PS\,\cite{Thi75}; TOFI-LANL\,\cite{Wou88};
KEK\,\cite{Koba91}; MSU\,\cite{Youn93};
MISTRAL-ISOLDE\,\cite{Bach08}; MAYA~\cite{Sav08} and TITAN [this work]. All shown with respect to the 2003 atomic mass evaluation~\cite{ame03}.  The second grey line shows the weighted average of all the values (which is essentially identical to the TITAN result). The three most recent results are shown inset on an expanded scale for better comparison.\label{s2n}
 }
\end{figure}

Using these mass measurements the two neutron separation energy, $S_{2n}$, of $^{11}$Li was  calculated to be $369.15(65)$ keV.  Figure~\ref{s2n} shows this new value
along with those calculated from all previous mass measurements of
$^{11}$Li.  The value from CERN-PS\,\cite{Thi75} was obtained using a magnetic dipole mass
spectrometer.  The TOFI-LANL\,\cite{Wou88} result is a time-of-flight
measurement of a fragmented beam using an isochronous mass
spectrometer.  The KEK\,\cite{Koba91} result is a
$^{11}$B($\pi^{-},\pi^{+}$)$^{11}$Li reaction $Q$-value and the
MSU\,\cite{Youn93} result is derived from the $Q$-value of the
$^{14}$C($^{11}$B,$^{11}$Li)$^{14}$O reaction.
The previous best result was achieved at ISOLDE by the
transmission spectrometer MISTRAL~\cite{Bach08}.  The MAYA experiment (also carried out at TRIUMF) used an active target to study the $^{11}$Li(p,t)$^{9}$Li reaction~\cite{Tani08}.  
The new $^9$Li value can be seen to be tens times more accurate than the literature value and both the values for $^{8,9}$Li show good agreement with previous measurements.
\begin{table}
\caption{\label{tab:meas} Frequency ratios, $r = \nu_{\rm{ref}} / \nu_{\rm{c}}$, for $^{8,9,11}$Li and the derived mass excesses, $\Delta$. Also shown are the literature values for the mass excesses for comparison~\cite{ame03}.  The $^8$Li literature value is derived by adding the average Q-value for the  $^7$Li(n, $\gamma$)$^8$Li reaction (as given in~\cite{ame03}) to the recent SMILETRAP measurement of the mass of $^7$Li~\cite{nag06}.}
\begin{ruledtabular}
\begin{tabular}{cccc}
Isotope & r  &    $\Delta_{\rm{TITAN}}$ (keV) &    $\Delta_{\rm{Lit}}$ (keV)  \\
%heading
\hline
$^8$Li &	1.333\,749\,862(18) &   20,945.80(11) & 20,945.799(65)\\
$^9$Li &	1.500\,728\,256(34) & 24,954.91(20) & 24,954.3(19) \\
$^{11}$Li & 1.836\,069\,26(11) & 40,728.28(64) & 40,797(19)\\
\end{tabular}
\end{ruledtabular}
\end{table}

Although in good agreement with the TOFI-LANL and KEK results the MISTRAL measurement shows over two sigma deviation from the MSU result.  Analysis of recent measurements of both the soft-dipole excitation, via invariant mass spectrometry, and the charge radius, via isotope shifts, of $^{11}$Li requires the mass.  However, due to this uncertainty in the mass the invariant mass spectrometry data were analysed using the AME03 value whereas the isotope shift measurements used the MISTRAL result.  It was reported in Nakamura  \textit{et al.} that using the MISTRAL result for the $^{11}$Li mass would enhance the total E1 strength by 6\%~\cite{Naka06}.  Using the AME mass
mass value for $^{11}$Li (11.043\,798(21) u) in the analysis of the isotope shift measurement results in a charge radius of 2.465(19) fm significantly
increased from  the value reported in Sanchez  \textit{et al}, 2.422(17) fm~\cite{San06}, which was  calculated using the MISTRAL value for the mass (11.043\,715\,7(54) u).  A good test for nuclear theory would be its ability to reproduce the results of both these experiments. However, a meaningful comparison is only possible if the two results are analysed using a consistent value for the $^{11}$Li mass.   With 1.5 sigma agreement with the MISTRAL result the new TITAN measurement shows that the MSU value for the  $^{11}$Li mass significantly under-binds the two neutron halo.  The impact of this new measurement is then clear. The results from the charge radius measurements and the soft-dipole excitation can now both be recalculated using the new mass. This 
will help to furnish a more consistent experimental picture of $^{11}$Li which will 
in turn prove an excellent test of nuclear theory. 

The nuclear charge radius obtained from laser spectroscopy of the 2s - 3s isotope shift
represents a remarkable confluence of correlation effects from
atomic theory on the one hand and nuclear theory on the other.  The
total isotope shift is the sum of a nuclear volume effect that is
(nearly) independent of nuclear mass, and a variety of atomic
structure effects that depend linearly and even quadratically on the
nuclear mass~\cite{yan08}.  Relativistic and quantum electrodynamic
corrections must be fully taken into account with electron
correlation in order to achieve sufficient accuracy in the atomic
structure part.  With a sufficiently accurate nuclear mass, the
atomic structure part can then be accurately calculated and
subtracted from the isotope shift in order to derive a value for the
nuclear charge radius~\cite{yan08}.  Using the TITAN value for the $^{11}$Li mass (11.043\,723\,61(69) u)
results in a nuclear charge radius of
2.427(16) fm.  This is a $2 \, \sigma$ deviation from the result calculated using the AME03 and the improved accuracy in the nuclear mass effectively
eliminates this as a source of uncertainty in the interpretation of
the isotope shift in terms of a nuclear charge radius, and so the
nuclear charge radius provides a rigorous test of neutron
correlations on the nuclear physics side.  (The residual uncertainty
of $\pm$0.016 fm comes from the uncertainty in the isotope shift
measurement itself.)  As discussed by Sanchez et al.~\cite{San06} and
Puchalski et al.~\cite{puc06}, the result shows that the $^9$Li core is
significantly perturbed by the presence of the two halo neutrons.

In summary we have performed a mass measurement on the exotic halo nucleus $^{11}$Li using the newly commissioned TITAN Penning trap mass spectrometer.  This measurement sets a new record for the shortest lived isotope ever measured using this technique with a half- life, 8.8 ms, over seven times shorter than the previous record ($^{74}$Rb, t$_{1/2}$ = 65 ms) held by ISOLTRAP.  Using the new measurements a value for the two neutron separation energy of $^{11}$Li was calculated, $S_{2n} = 369.15(65)$ keV with a precision over seven times better than the best previous result.

\begin{acknowledgments}
This work has been partially supported by the Natural
Sciences and Engineering Research Council of Canada
(NSERC). TRIUMF receives federal funding via a contribution
agreement with the National Research Council of
Canada (NRC).
\end{acknowledgments}

\end{document}